# On the angular distribution and spin polarization of the photoelectrons from semi-filled shell atoms

**M. Ya. Amusia** [a,b)] **and L. V. Chernysheva** [b)],

[a)] *Racah Institute of Physics, The Hebrew University, Jerusalem 91904, Israel*
[b)] *A. F. Ioffe Physical- Technical Institute, St. Petersburg 194021, Russia*

*Key words*: Spin-polarization parameters, intro-doublet interaction, electron correlations

**Abstract**

We present here the results of calculations of photoelectrons' angular anisotropy and spin-polarization parameters for a number of semi-filled shell atoms. We consider ionization of outer or in some cases next to the outer electrons in a number of elements from I, V, and VI groups of the Periodic Table. All calculations are performed with account of multi-electron correlations in the frame of the Spin Polarized version of the Random Phase Approximation with Exchange – SP RPAE. We consider the dipole angular distribution and spin polarization of photoelectrons from semi-filled subshells and from closed shells that are neighbours to the semi-filled shells.

We have considered also angular anisotropy and spin-polarization of photoelectrons from some excited atoms that are formed by spin-flip of one of the outer electrons. To check the accuracy and consistency of the applied SP RPAE approach and to see the role of the nuclear charge variation only, we have calculated the dipole angular anisotropy and spin-polarization parameters of $3p$ - electrons in K and compare them to Ar and $K^+$ that have the same configuration.

Entirely, we have calculated the angular anisotropy and spin-polarization parameters for following subshells of atoms N ($2p$), P ($3p$), Ar ($3p$), $K^+$($3p$), K($3p$), Cr($3p$, $3d$), $Cr^*$($3d$), Mn($3p$, $3d$), As($3d$, $4p$), Mo($4p$, $4d$), $Mo^*$($4d$), Tc($4p$, $4d$, ), Sb($4d$, $5p$), Eu($4f$).

The peculiarities of obtained parameters as function of photon frequencies are discussed, as well as some specific features of considered semi-filled shell objects.

PACS: 32.80.-t, 32.80.Fb, 33.60.-q, 33.80.Eh

## 1. Introduction

Recently we performed calculations of dipole and non-dipole spin-polarization parameters of photoelectrons from Xe, Cs, and Ba $3d$ spin-orbit components of $3d_{5/2}$ and $3d_{3/2}$ doublet [1, 2]. We have treated there the components of spin-orbit doublet as semi-filled subshells.

However, there exists a number of atoms, almost one-third of the Periodic Table, that can be considered as semi-filled shell objects and treated almost as accurate as close shell atoms, accounting for electron correlations in the frame of so-called Spin-Polarised Random Phase Approximation with Exchange – SP RPAE [3, 4]. In the frame of this approximation we studied recently the non-dipole angular anisotropy parameters of semi-filled shell atoms [5]. These parameters were calculated long before for Mn only [6].

Angular distribution in the dipole approximation has been studied for a number of closed shell atoms (see e.g. [3, 4]). Attention was given also to semi-filled shell atoms Cr and Mn [7, 8].

Spin polarization parameters that determine the orientation of photoelectron spin relative to the photons flux or to their polarization direction are important characteristics of the photoionization process. It was demonstrated long ago [9], that spin polarization of photoelectrons, although it is caused by relatively weak spin-orbit interaction, is not a small relativistic effect. It appeared that the spin-orbit interaction acts as a leverage that permits to disclose the degree of photoelectrons spin polarization that can reach at some emission angles 100% [10].

At the same time, there was a belief that investigation of spin polarization of photoelectrons permits to realize the so-called complete experiment in atomic physics, since the five parameters, the partial photoionization cross section $\sigma_{nl}(\omega)$ of the $nl$-subshell, photoelectron's angular anisotropy parameter $\beta_{nl}(\omega)$, and three characteristics of photoelectron's spin polarization $A_{nl}^j(\omega)$, $\alpha_{nl}^j(\omega)$ and $\xi_{nl}^j(\omega)$, where $j$ is the total momentum of the ionized shell, are sufficient to determine completely five theoretical values, three photoionization amplitudes $D_{nl\to\varepsilon(l\pm 1)}^j(\omega)$ and two phase shift differences [11].

Later it was found that these parameters are not mutually independent, and there is one equation connecting them [12]. Therefore it is not possible to determine from the experiment of that kind 5 theoretical values. Still it is possible to determine in that way three non-relativistic values, two amplitudes and one phase shift difference. As the latest experiment shows, the non-relativistic approximation is quite sufficient even in such a heavy atom as Xe [12], and demonstrates the ability of this approach to describe satisfactorily the photoionization process. Different calculations demonstrated an important role played by multi-electron correlations in photoionization and proved that the latter can be accurately enough taken into account in the frames of the Random Phase Approximation with Exchange (RPAE) and some of its straightforwardly generalized versions, e.g. GRPAE or, for semi-filled shell atoms, SP RPAE [3, 4].

In this paper for the first time we have performed systematic studies of dipole angular anisotropy and spin-polarization parameters of photoelectrons from semi-filled shell atoms. The calculations will be performed in the frame of SP RPAE thus including the effect of important inter-electron correlations. We are expecting that new specific features will appear in the angular anisotropy and spin-polarization parameters.

**2. Equations for spin polarization**

Here we follow the paper [1], while more details can be found in [10, 11]. The formula for the photoelectron flux $I_{JJ'}(\vec{x},\vec{s})$, that originates from photoionization of an atom with initial total angular momentum $J$ and final state ion momentum $J'$, and for corresponding polarization parameters $A_{nl}^j(\omega)$, $\alpha_{nl}^j(\omega)$ and $\xi_{nl}^j(\omega)$ were derived in [10, 11]. For circularly polarized light one has [11]

$$I_{JJ'}(\omega) = \sum_j \frac{\sigma_{njl}}{8\pi} \left\{ 1 - \frac{1}{2}\beta_{nl}(\omega)\left[\frac{3}{2}(\vec{x}\vec{s})^2 - \frac{1}{2}\right] + A_{nl}^j(\omega)(\vec{s}\vec{s}_\gamma) - \right.$$
$$\left. - \alpha_{nl}^j(\omega)\left[\frac{3}{2}(\vec{x}\vec{s}_\gamma)(\vec{x}\vec{s}) - \frac{1}{2}(\vec{s}\vec{s}_\gamma)\right] + \xi_{nl}^j(\omega)[\vec{s}(\vec{x}\times \vec{k}_\gamma)(\vec{x}\vec{k}_\gamma)]\right\}$$
(1)

Here $s_\gamma$ and $s$ are unit vectors in photon and photoelectron spin directions, **x** is the direction of photoelectron's motion and $k_\gamma$ is the unit vector in the direction of photon motion. The photoelectron total angular momentum $j$ can take two values, $j = l \pm 1/2$. Similar expression for linearly polarized light can be found in [3].

The parameters $A_{nl}^j(\omega)$, $\alpha_{nl}^j(\omega)$ and $\xi_{nl}^j(\omega)$ are given by the following relations:

$$\beta_{nl}(\omega) = \frac{1}{2l+1}[(l+2)d_{l+1}^2 + (l-1)d_{l-1}^2 + 6\sqrt{l(l+1)}d_{l-1}d_{l+1}\cos(\delta_{l+1} - \delta_{l-1})] \times [d_{l+1}^2 + d_{l-1}^2]^{-1},$$

$$A_{nl}^j(\omega) = \frac{(-1)^{j-l-1/2}}{2j+1} \frac{ld_{l+1}^2 - (l+1)d_{l-1}^2}{d_{l-1}^2 + d_{l+1}^2},$$

$$\alpha_{nl}^j(\omega) = \frac{2(-1)^{j-l-1/2}l(l+1)}{(2j+1)(2l+1)}[(l+2)ld_{l+1}^2 - (l-1)(l+1)d_{l-1}^2 - 3\sqrt{l(l+1)}d_{l+1}d_{l-1}\cos(\delta_{l+1} - \delta_{l-1})]$$

$$\times [d_{l-1}^2 + d_{l+1}^2]^{-1},$$

$$\xi_{nl}^j = \frac{3(-1)^{j-l-1/2}\sqrt{l(l+1)}}{2j+1} \frac{d_{l+1}d_{l-1}\sin(\delta_{l+1} - \delta_{l-1})}{d_{l-1}^2 + d_{l+1}^2}.$$
(2)

Here

$$d_{l\pm 1} \equiv d_{nl,\varepsilon l\pm 1} = (-1)^{l_>}\sqrt{l_>}\int_0^\infty \phi_{nl}(r) r \phi_{\varepsilon l\pm 1}(r) dr,$$
(3)

with $l_> = l+1$ and $l_> = l$ for $l \to l+1$, $l \to l-1$ transitions, respectively, while $\phi_{nl}(r)$ and $\phi_{\varepsilon l'}(r)$ are the radial parts of the Hartree-Fock one-electron wave functions; $\delta_{l\pm 1}$ are the photoelectron $l \pm 1$ waves scattering phases.

The partial cross-section reaches its minimal value if at some $\omega$ the amplitude $d_{l+1}$ is equal to zero. At this point the parameters (2) are

$$\beta_{nl}(\omega) = \frac{l-1}{2l+1}; A_{nl}^j(\omega) = -\frac{(-1)^{j-l-1/2}(l+1)}{2j+1}; \alpha_{nl}^j(\omega) = -\frac{2(-1)^{j-l-1/2}(l-1)l(l+1)^2}{(2j+1)(2l+1)},$$
$$\xi_{nl}^j = 0.$$
(4)

In RPAE or GRPAE the parameters $A_{nl}^j(\omega)$, $\alpha_{nl}^j(\omega)$ and $\xi_{nl}^j(\omega)$ can be obtained using the following substitutions from [5]:

$$d_{l\pm 1}^2 \to |D_{l\pm 1}|^2,$$
$$d_{l+1}d_{l-1}\cos(\delta_{l+1} - \delta_{l-1}) \to \text{Re}[D_{l+1}D_{l-1}^* e^{i(\delta_{l+1} - \delta_{l-1})}],$$
$$d_{l+1}d_{l-1}\sin(\delta_{l+1} - \delta_{l-1}) \to \text{Im}[D_{l+1}D_{l-1}^* e^{i(\delta_{l+1} - \delta_{l-1})}].$$
(5)

As a result, one has

$$\beta_{nl}(\omega) = \frac{1}{(2l+1)[|D_{l-1}|^2 + |D_{l+1}|^2]}[(l+2)|D_{l+1}|^2 - (l-1)|D_{l-1}|^2 +$$
$$+ 6\sqrt{l(l+1)}[(\operatorname{Re} D_{l+1} \operatorname{Re} D_{l-1} + \operatorname{Im} D_{l+1} \operatorname{Im} D_{l-1})\cos(\delta_{l+1} - \delta_{l-1}) - \qquad (6)$$
$$- (\operatorname{Re} D_{l-1} \operatorname{Im} D_{l+1} - \operatorname{Re} D_{l+1} \operatorname{Im} D_{l-1}]\sin(\delta_{l+1} - \delta_{l-1})],$$

$$A_{nl}^{j}(\omega) = \frac{(-1)^{j-l-1/2} l(l+1)}{2j+1} \frac{|D_{l+1}|^2 - |D_{l+1}|^2}{|D_{l-1}|^2 + |D_{l+1}|^2}, \qquad (7)$$

$$\alpha_{nl}^{j}(\omega) = \frac{2(-1)^{j-l-1/2} l(l+1)}{(2j+1)(2l+1)[|D_{l-1}|^2 + |D_{l+1}|^2]}\{(l+2)l|D_{l+1}|^2 - (l-1)(l+1)|D_{l-1}|^2 -$$
$$- 3\sqrt{l(l+1)}[(\operatorname{Re} D_{l+1} \operatorname{Re} D_{l-1} + \operatorname{Im} D_{l+1} \operatorname{Im} D_{l-1})\cos(\delta_{l+1} - \delta_{l-1}) - \qquad (8)$$
$$- (\operatorname{Re} D_{l-1} \operatorname{Im} D_{l+1} - \operatorname{Re} D_{l+1} \operatorname{Im} D_{l-1}]\sin(\delta_{l+1} - \delta_{l-1})\},$$

$$\xi_{nl}^{j} = \frac{3(-1)^{j-l-1/2}\sqrt{l(l+1)}}{2j+1} \frac{1}{d_{l-1}^2 + d_{l+1}^2}[(\operatorname{Re} D_{l+1} \operatorname{Re} D_{l-1} + \operatorname{Im} D_{l+1} \operatorname{Im} D_{l-1})\sin(\delta_{l+1} - \delta_{l-1}) + \qquad (9)$$
$$+ (\operatorname{Re} D_{l-1} \operatorname{Im} D_{l+1} - \operatorname{Re} D_{l+1} \operatorname{Im} D_{l-1}]\cos(\delta_{l+1} - \delta_{l-1})],$$

In this paper we intend to study the angular anisotropy and spin polarization parameters for $p$-, $d$-, and $f$-electrons, i.e. for $l=1, 2. 3$. It is seen from (7, 8, 9), that $A_{nl}^{3/2}(\omega)$, $\alpha_{nl}^{3/2}(\omega)$, $\xi_{nl}^{3/2}(\omega)$ and $A_{nl}^{1/2}(\omega)$, $\alpha_{nl}^{1/2}(\omega)$, $\xi_{nl}^{1/2}(\omega)$ for $l=1$, $A_{nl}^{5/2}(\omega)$, $\alpha_{nl}^{5/2}(\omega)$, $\xi_{nl}^{5/2}(\omega)$ and $A_{nl}^{3/2}(\omega)$, $\alpha_{nl}^{3/2}(\omega)$, $\xi_{nl}^{3/2}(\omega)$ for $l=2$, $A_{nl}^{7/2}(\omega)$, $\alpha_{nl}^{7/2}(\omega)$, $\xi_{nl}^{7/2}(\omega)$ and $A_{nl}^{5/2}(\omega)$, $\alpha_{nl}^{5/2}(\omega)$, $\xi_{nl}^{5/2}(\omega)$ for $l=3$ differ, respectively, only by their signs and factors $(2j+1)$.

### 3. Inclusion of intra-doublet correlations

To perform calculations in one-electron HF approximation, the matrix elements (5) and phases were calculated using computer codes described in [4].

Atoms with semi-filled shells can be treated in the frame of the frame SP RPAE. The SP RPAE equations are rather complex and can be found in [3] or [4], so we present them below in the symbolic form. The symbolic version of the equations is as follows

$$(D_\uparrow(\omega)\ D_\downarrow(\omega)) = (d_\uparrow\ d_\downarrow) + (D_\uparrow(\omega)\ D_\downarrow(\omega))\begin{pmatrix} \chi_{\uparrow\uparrow} & 0 \\ 0 & \chi_{\downarrow\downarrow} \end{pmatrix}\begin{pmatrix} U_{\uparrow\uparrow} & V_{\uparrow\downarrow} \\ V_{\downarrow\uparrow} & U_{\downarrow\downarrow} \end{pmatrix}. \qquad (10)$$

Here signs ↑ and ↓ denote the "up" and "down" photoelectron vacancy spin projection, $U_{\uparrow\uparrow(\downarrow\downarrow)}$ are the combinations of the direct and exchange Coulomb interelectron interaction, $U = V^d - V^e$ matrix elements, while $V_{\uparrow\uparrow(\downarrow\downarrow)}$ are the pure Coulomb matrix elements.

By solving (10) we will concentrate on the investigation of the influence of "up" and "down" electrons upon themselves and each other.

## 4. Results of calculations and discussion

Entirely, we have calculated the angular anisotropy and spin-polarization parameters for the following subshells of atoms N (2$p$), P (3$p$), Ar (3$p$), K$^+$(3$p$), K(3$p$), Cr(3$p$, 3$d$), Cr$^*$(3$d$), Mn(3$p$, 3$d$), As(3$d$, 4$p$), Mo(4$p$, 4$d$), Mo$^*$(4$d$), Tc(4$p$, 4$d$, ), Sb(4$d$, 5$p$), Eu(4$f$).

Due to interaction with the semi-filled subshell electrons each closed subshell splits into two "up" and "down" levels denoted as ↑ or ↓, according to the individual electrons spin projection. Thus, the electron configuration of the considered objects is presented in the following way:

$N 1s \uparrow 1s \downarrow 2s \uparrow 2s \downarrow \underline{2p^3} \uparrow$

$P 1s \uparrow 1s \downarrow 2s \uparrow 2s \downarrow 2p^3 \uparrow 2p^3 \downarrow 3s \uparrow 3s \downarrow 3p^3 \uparrow \equiv [Ne \uparrow\downarrow] 3s \uparrow 3s \downarrow \underline{3p^3} \uparrow$

$Ar, K^+ 1s^2 2s^2 2p^6 3s^2 3p^6$

$K 1s \uparrow 1s \downarrow 2s \uparrow 2s \downarrow 2p^3 \uparrow 2p^3 \downarrow 3s \uparrow 3s \downarrow 3p^3 \uparrow 3p^3 \downarrow 4s \uparrow \equiv [Ar \uparrow\downarrow] \underline{4s} \uparrow$

$Cr [Ar \uparrow\downarrow] \underline{3d^5} \uparrow \underline{4s} \uparrow$

$Cr^* [Ar \uparrow\downarrow] \underline{3d^5} \uparrow \underline{4s} \downarrow$

$Mn [Ar \uparrow\downarrow] \underline{3d^5} \uparrow 4s \uparrow 4s \downarrow$

$As [Ar \uparrow\downarrow] \underline{3d^5} \uparrow \underline{3d^5} \downarrow 4s \uparrow 4s \downarrow \underline{4p^3} \uparrow$

$Mo [Kr \uparrow\downarrow] \underline{4d^5} \uparrow \underline{5s} \uparrow$

$Mo^* [Kr \uparrow\downarrow] \underline{4d^5} \uparrow \underline{5s} \downarrow$

$Tc [Kr \uparrow\downarrow] \underline{4d^5} \uparrow 5s \uparrow 5s \downarrow$

$Sb [Kr \uparrow\downarrow] \underline{4d^5} \uparrow \underline{4d^5} \downarrow 5s \uparrow 5s \downarrow \underline{5p} \uparrow$

$Eu [Xe \uparrow\downarrow] \underline{4f^7} 6s \uparrow 6s \downarrow$

Single-underlined are the semi-filled subshells. Double-underlined are the considered *up* and *down* levels formed from closed subshells formed due to presence of a semi-filled subshell, for which the spin-polarization parameters are considered. The notation $[Ar \uparrow\downarrow, Kr \uparrow\downarrow, Xe \uparrow\downarrow]$ means that angular anisotropy and spin-polarization parameters were calculated for $3p \uparrow 3p \downarrow; 4p \uparrow 4p \downarrow; 5p \uparrow 5p \downarrow$ electrons belonging to the Ar-, Kr- and Xe-type core, respectively. We have calculated the angular anisotropy and spin-polarization parameters for single-underlined $p, d, f$ - semi-filled subshells. For s-electrons the angular anisotropy parameter is equal to 2 [see (6)] while spin-polarization parameters are according to equations (7)-(9) equal to zero.

The results of calculations are presented in Figures 1 -16. We see that all the parameters are rather complicated functions of frequency. Particularly strong is the variation of parameters in the vicinity of thresholds, where this is caused to large extend by rather fast variation of the phase differences and their sine's and cosine's functions that enter the angular anisotropy and spin-polarization parameters (6-9).

The values of all parameters at the cross section minimum are given, without SP RPAE corrections, by (4).

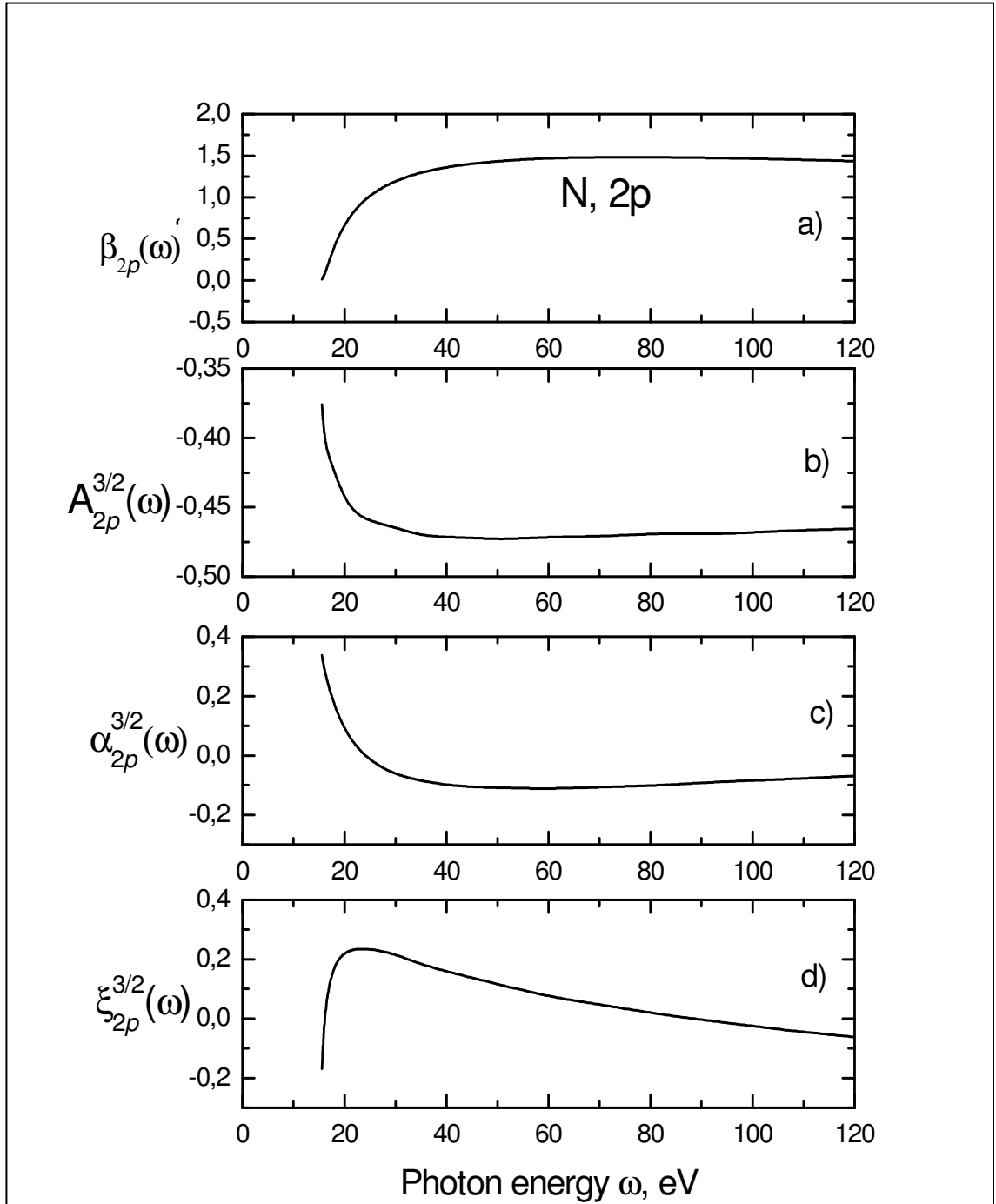

Fig. 1. Dipole anisotropy parameters of photoelectrons from N $2p^3$-subshell as functions of photon energy $\omega$:
a) angular $\beta_{2p}(\omega)$; spin-polarization b) $A_{2p}^{3/2}(\omega)$, c) $\alpha_{2p}^{3/2}(\omega)$, d) $\xi_{2p}^{3/2}(\omega)$ parameters.

It is seen that all variations are concentrated in the near threshold region.

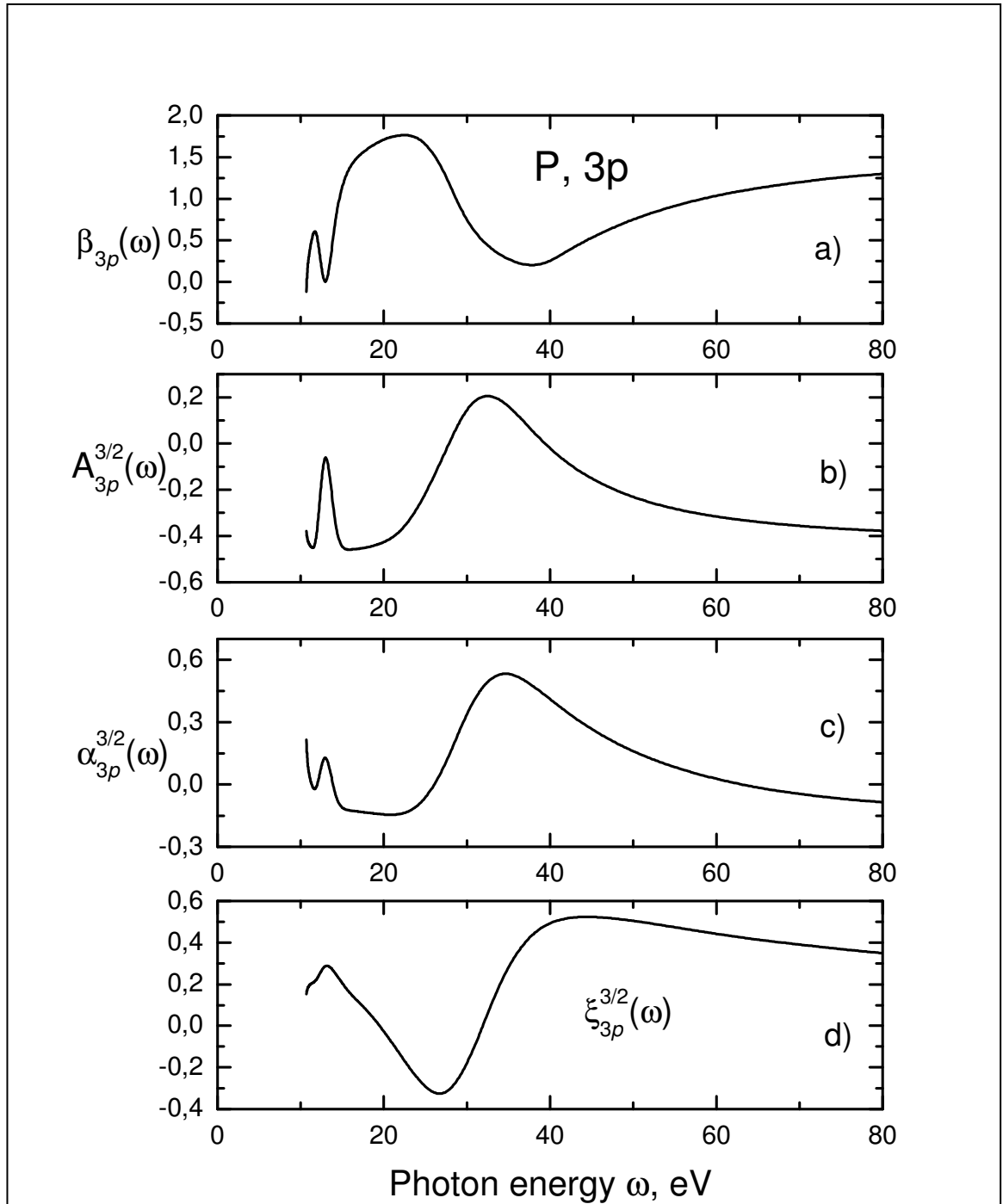

Fig. 2. Dipole anisotropy parameters of photoelectrons from P $3p^3$-subshell as functions of photon energy $\omega$:

a) angular $\beta_{3p}(\omega)$; spin-polarization b) $A^{3/2}_{3p}(\omega)$, c) $\alpha^{3/2}_{3p}(\omega)$, d) $\xi^{3/2}_{3p}(\omega)$ parameters.

Here the curves are much more complex than in case of N, having two powerful maxima and a peculiar structure close to threshold.

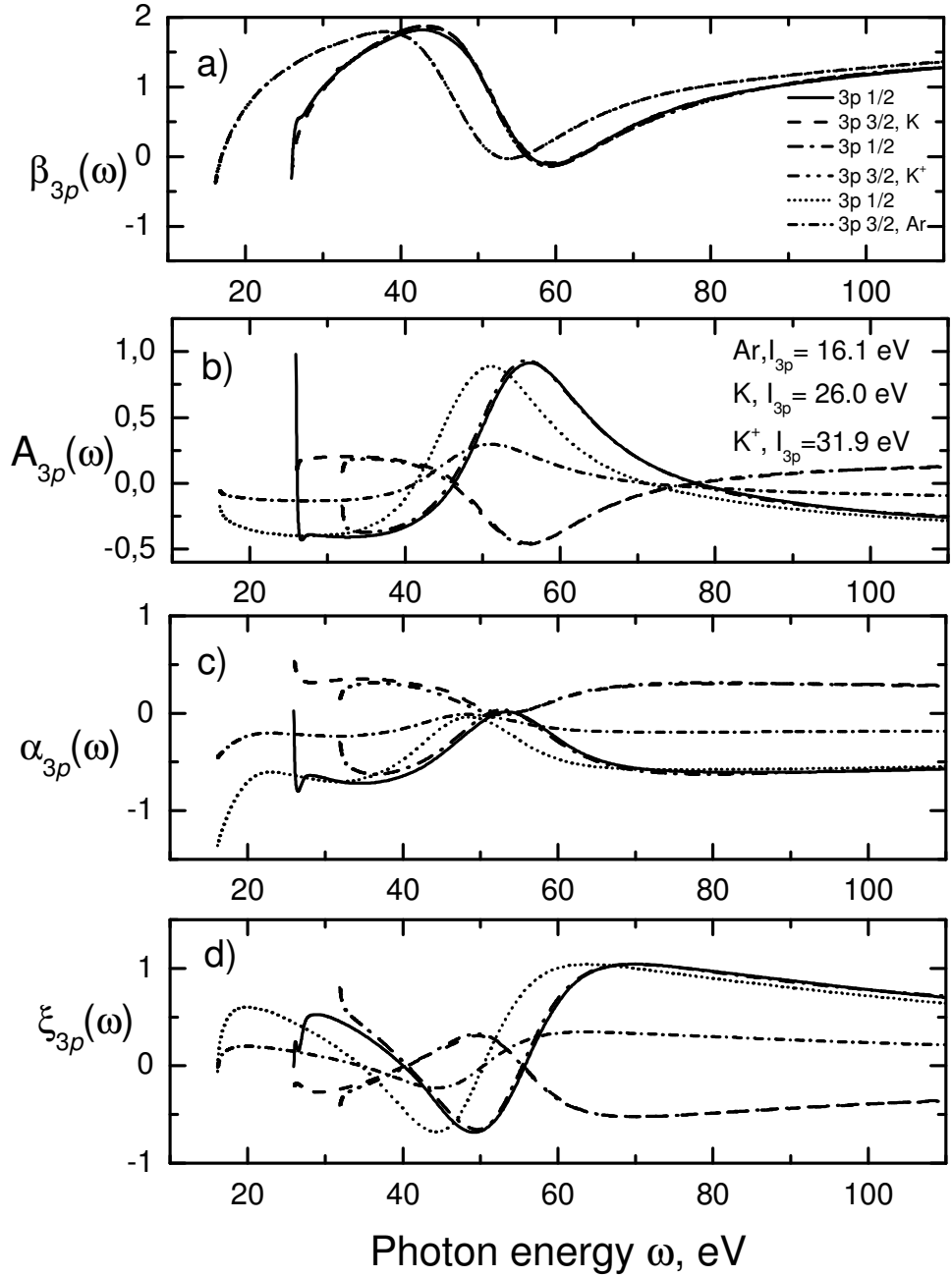

Fig. 3. Dipole anisotropy parameters of photoelectrons from Ar, K, K$^+$ 3p$^3$-subshell as functions of photon energy $\omega$:
a) angular $\beta_{3p}(\omega)$; spin-polarization b) $A_{3p}(\omega)$, c) $\alpha_{3p}(\omega)$, d) $\xi_{3p}(\omega)$ parameters.

The changes that take place on the way from Ar to K are smooth. Only in K there are strong near-threshold variation. Of interest is the parameters variation at 40 – 60 eV.

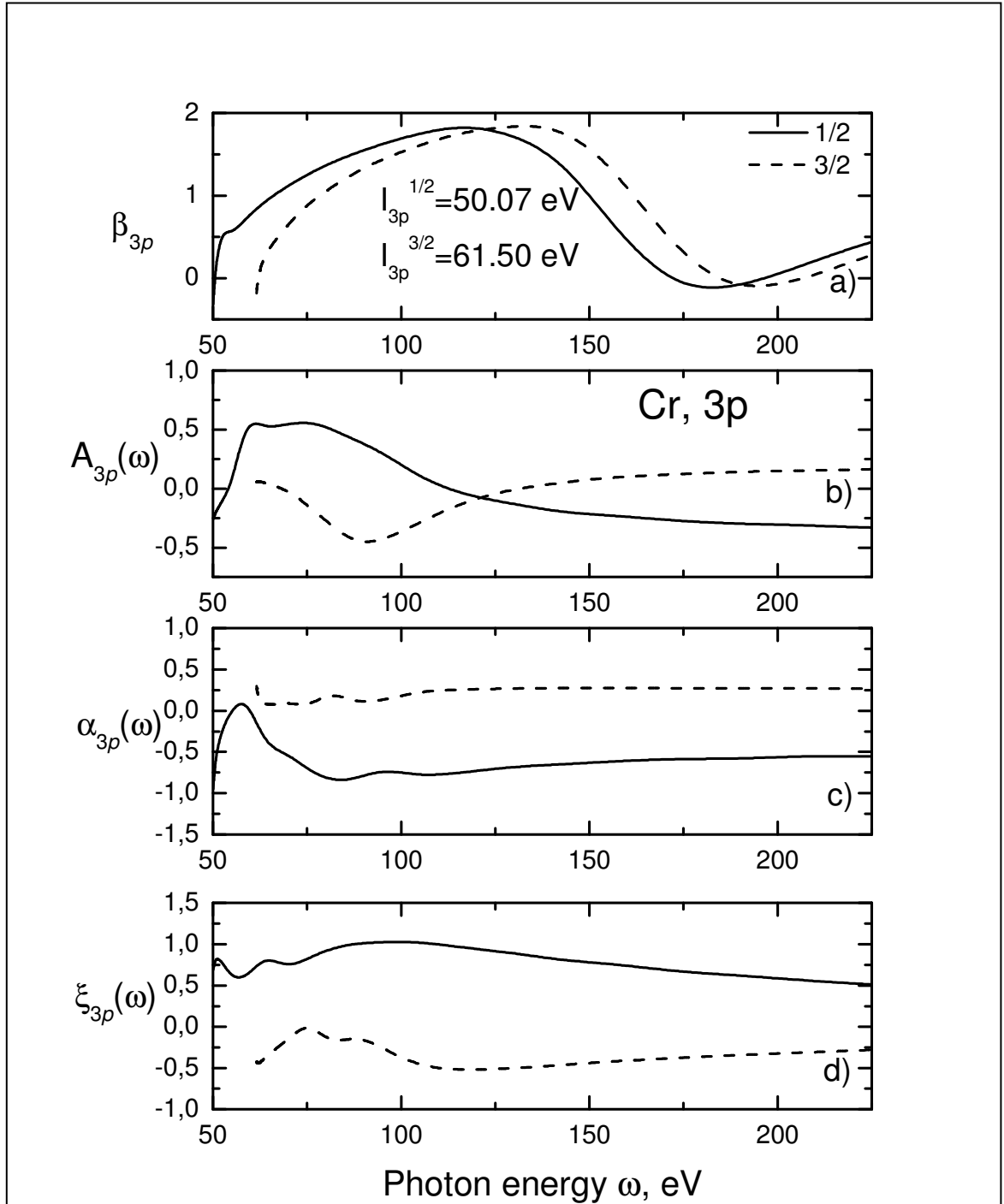

Fig. 4. Dipole anisotropy parameters of photoelectrons from Cr $3p^3$-subshell as functions of photon energy $\omega$:
a) angular $\beta_{3p}(\omega)$; spin-polarization b) $A_{3p}(\omega)$, c) $\alpha_{3p}(\omega)$, d) $\xi_{3p}(\omega)$ parameters.

This figure demonstrates a prominent difference in spin-polarization parameters for *up* and *down* 3p-levels in Cr, thus illustrating the effect of $3d^5$ *up* electrons in Cr.

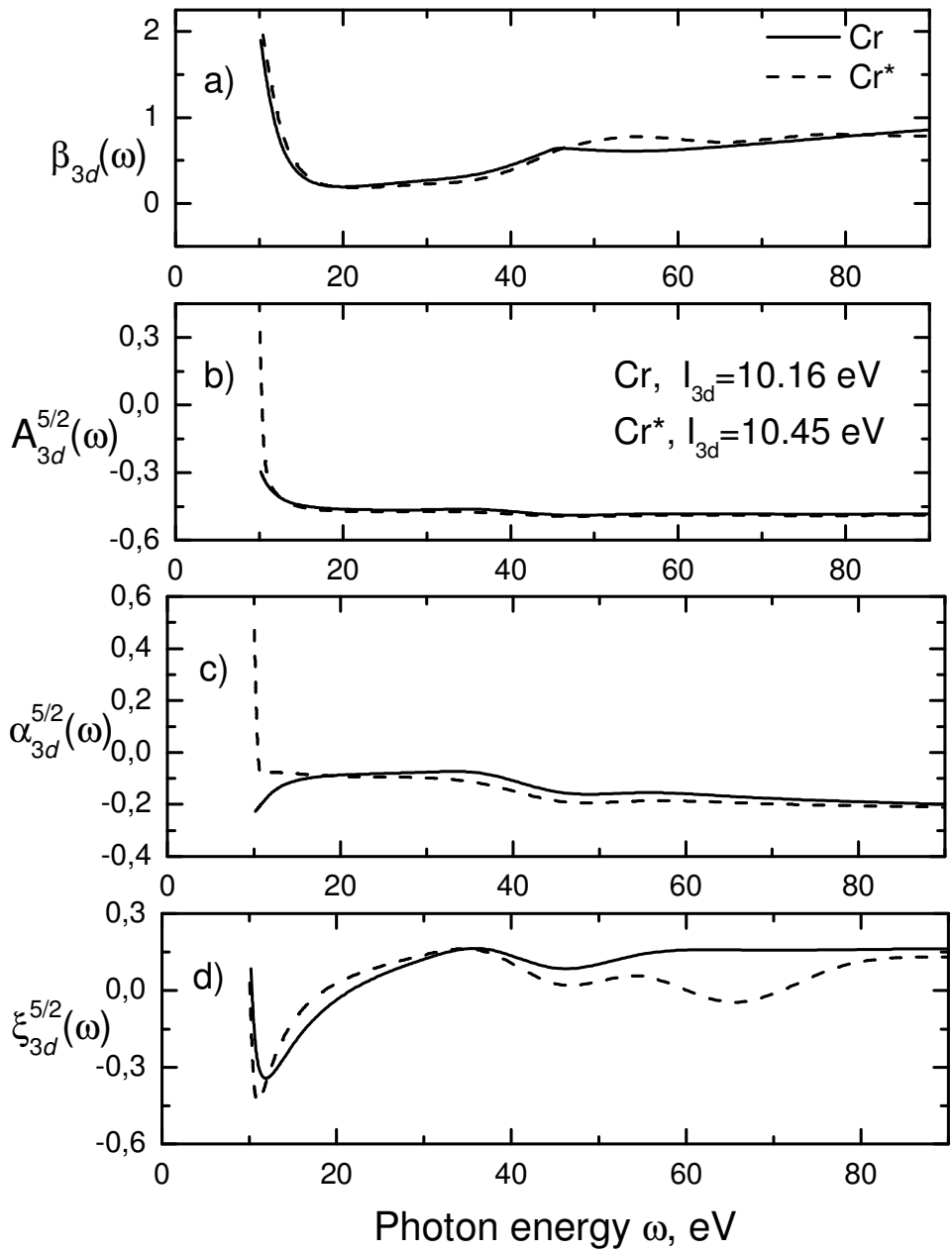

Fig. 5. Dipole anisotropy parameters of photoelectrons from Cr and Cr* $3d^5$-subshell as functions of photon energy $\omega$:
a) angular $\beta_{3d}(\omega)$; spin-polarization b) $A_{dn}^{5/2}(\omega)$, c) $\alpha_{3d}^{5/2}(\omega)$, d) $\xi_{3d}^{5/2}(\omega)$ .parameters.

It is seen, that the variation of the outer electron 4s spin projection affects the considered parameters rather weak, except very close to threshold for A and $\alpha$.

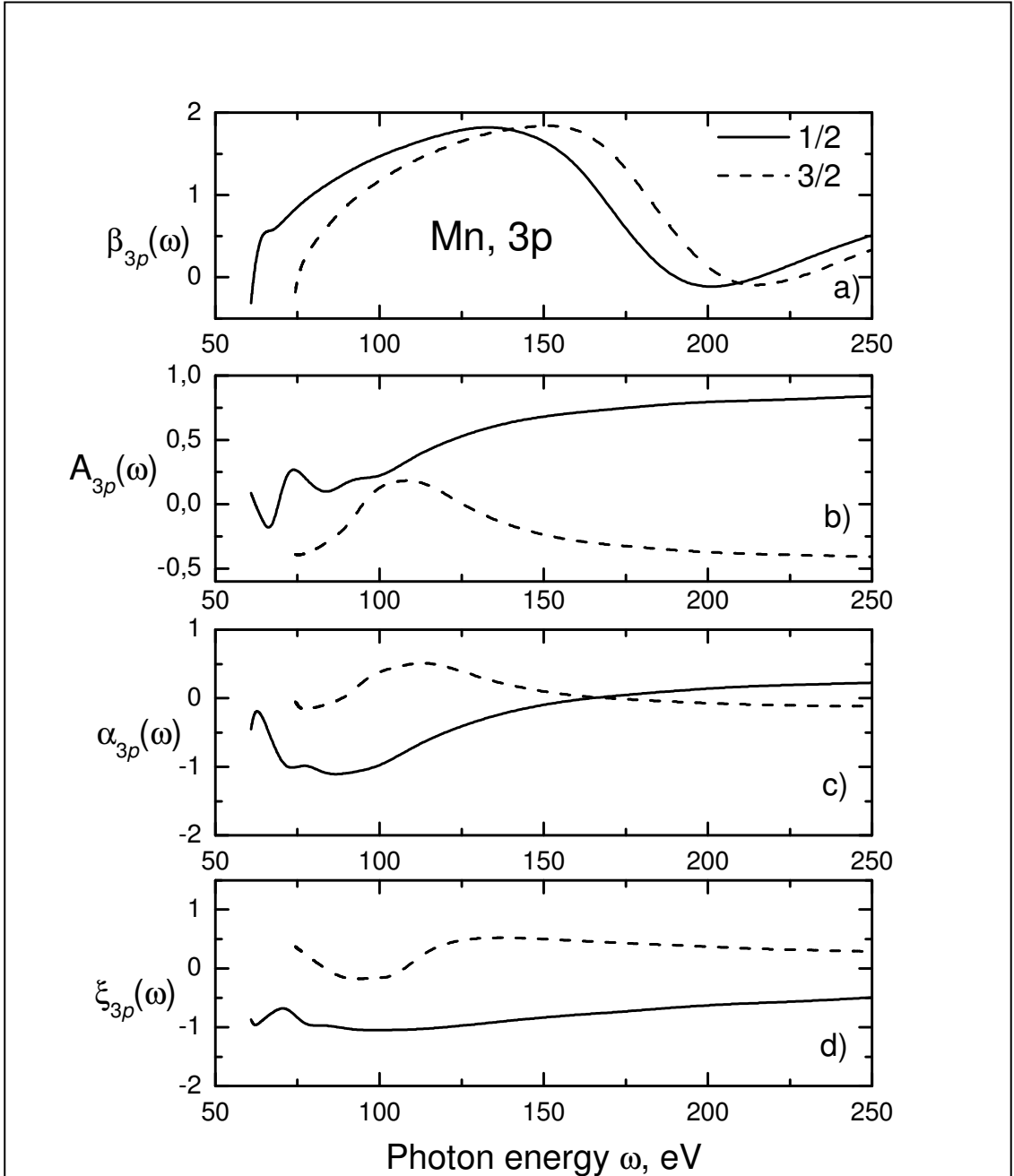

Fig. 6. . Dipole anisotropy parameters of photoelectrons from Mn $3p^3$-subshell as functions of photon energy $\omega$:
a) angular $\beta_{3p}(\omega)$; spin-polarization b) $A_{3p}(\omega)$, c) $\alpha_{3p}(\omega)$, d) $\xi_{3p}(\omega)$ parameters.

While $\beta$ parameters are similar for Mn and Cr, their spin-polarization parameters have almost nothing in common, both in magnitude and in shape. This demonstrates the sensitivity of thee parameters upon atomic structure.

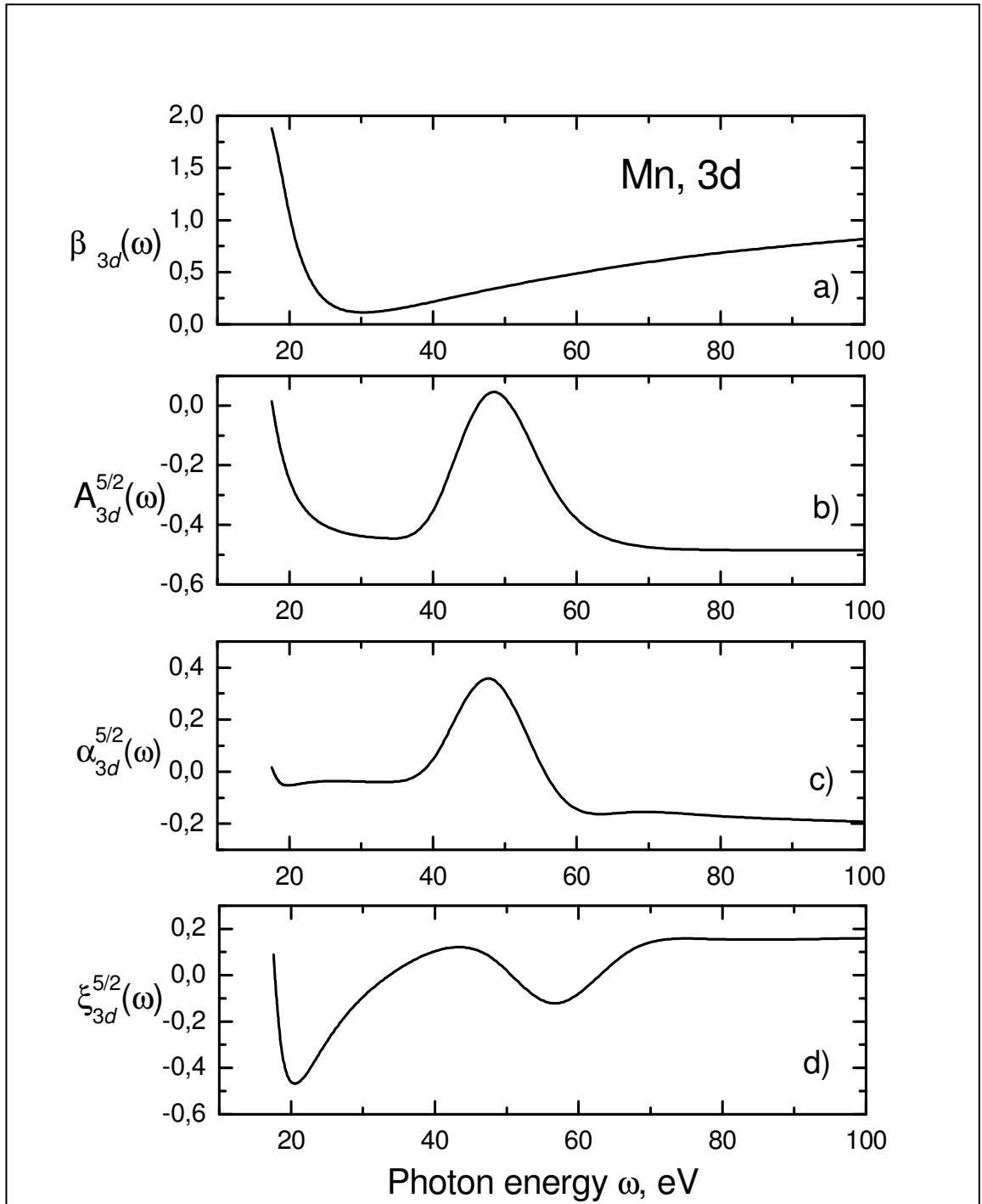

Fig. 7. Dipole anisotropy parameters of photoelectrons from Mn $3d^5$-subshell as functions of photon energy $\omega$:
a) angular $\beta_{3d}(\omega)$; spin-polarization b) $A_{3d}^{5/2}(\omega)$, c) $\alpha_{3d}^{5/2}(\omega)$, d) $\xi_{3d}^{5/2}(\omega)$ parameters.

Of particular interest is the maximum in A and $\alpha$ at 50 eV and prominent variation in $\xi$ that does not exist in Cr.

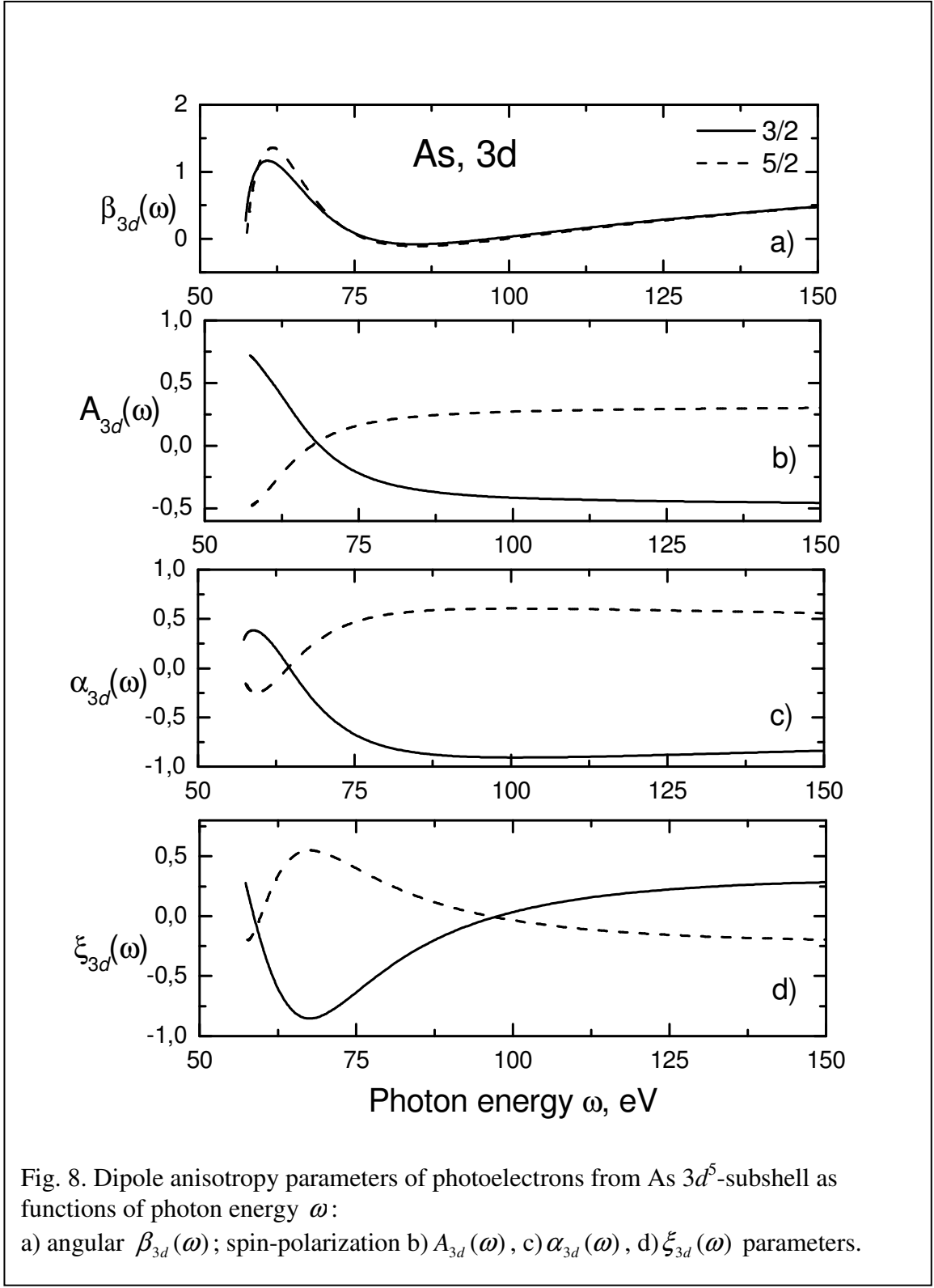

Fig. 8. Dipole anisotropy parameters of photoelectrons from As $3d^5$-subshell as functions of photon energy $\omega$:
a) angular $\beta_{3d}(\omega)$; spin-polarization b) $A_{3d}(\omega)$, c) $\alpha_{3d}(\omega)$, d) $\xi_{3d}(\omega)$ parameters.

As one could assume, the outer semi-filled subshell $3p^3$ almost do not affect the considered parameters for the inner $3d^5$ electrons.

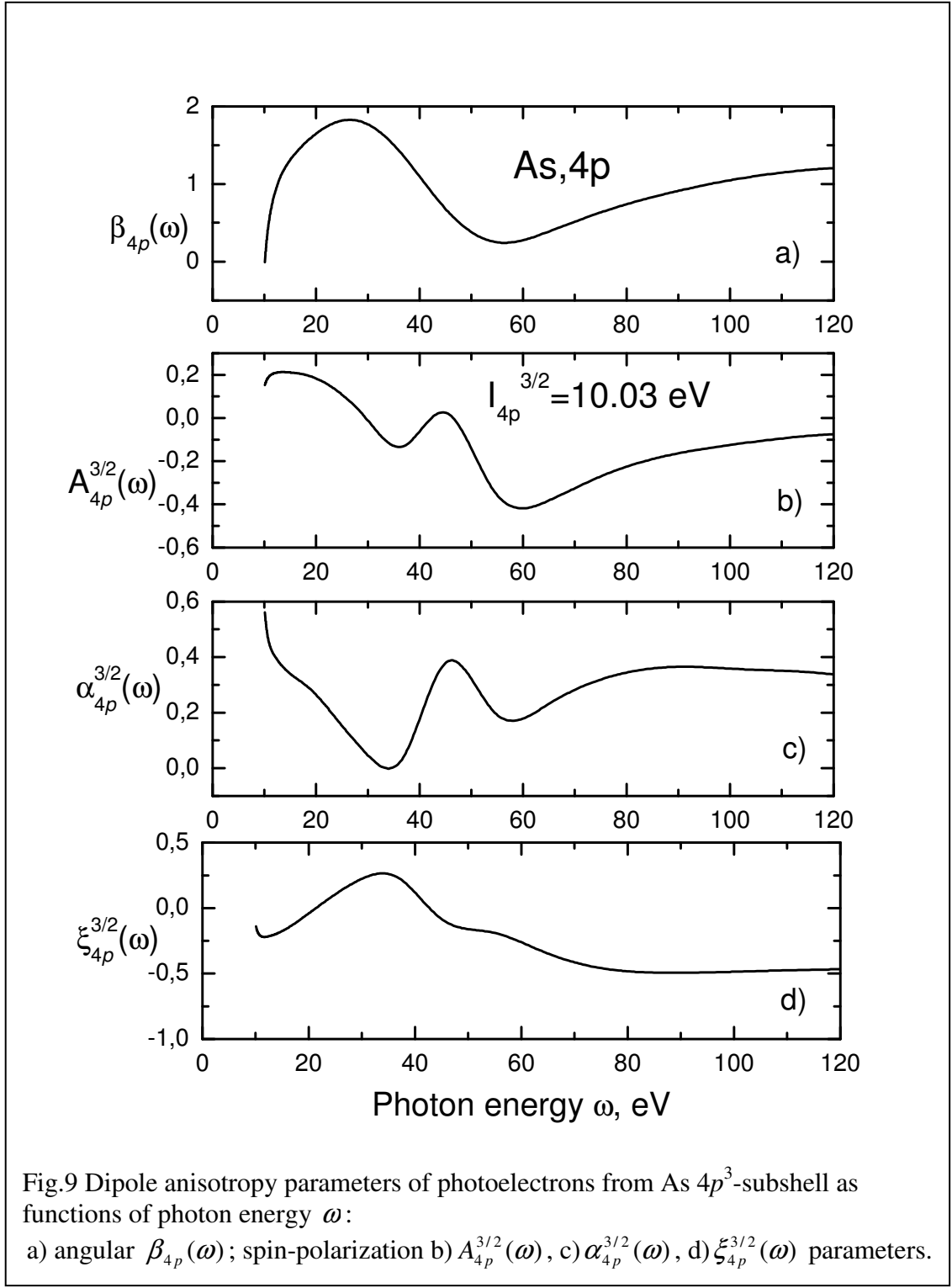

Fig.9 Dipole anisotropy parameters of photoelectrons from As $4p^3$-subshell as functions of photon energy $\omega$:
a) angular $\beta_{4p}(\omega)$; spin-polarization b) $A_{4p}^{3/2}(\omega)$, c) $\alpha_{4p}^{3/2}(\omega)$, d) $\xi_{4p}^{3/2}(\omega)$ parameters.

While $\beta$ for As is similar to that of the same group element P, the spin-polarization parameters are essentially different.

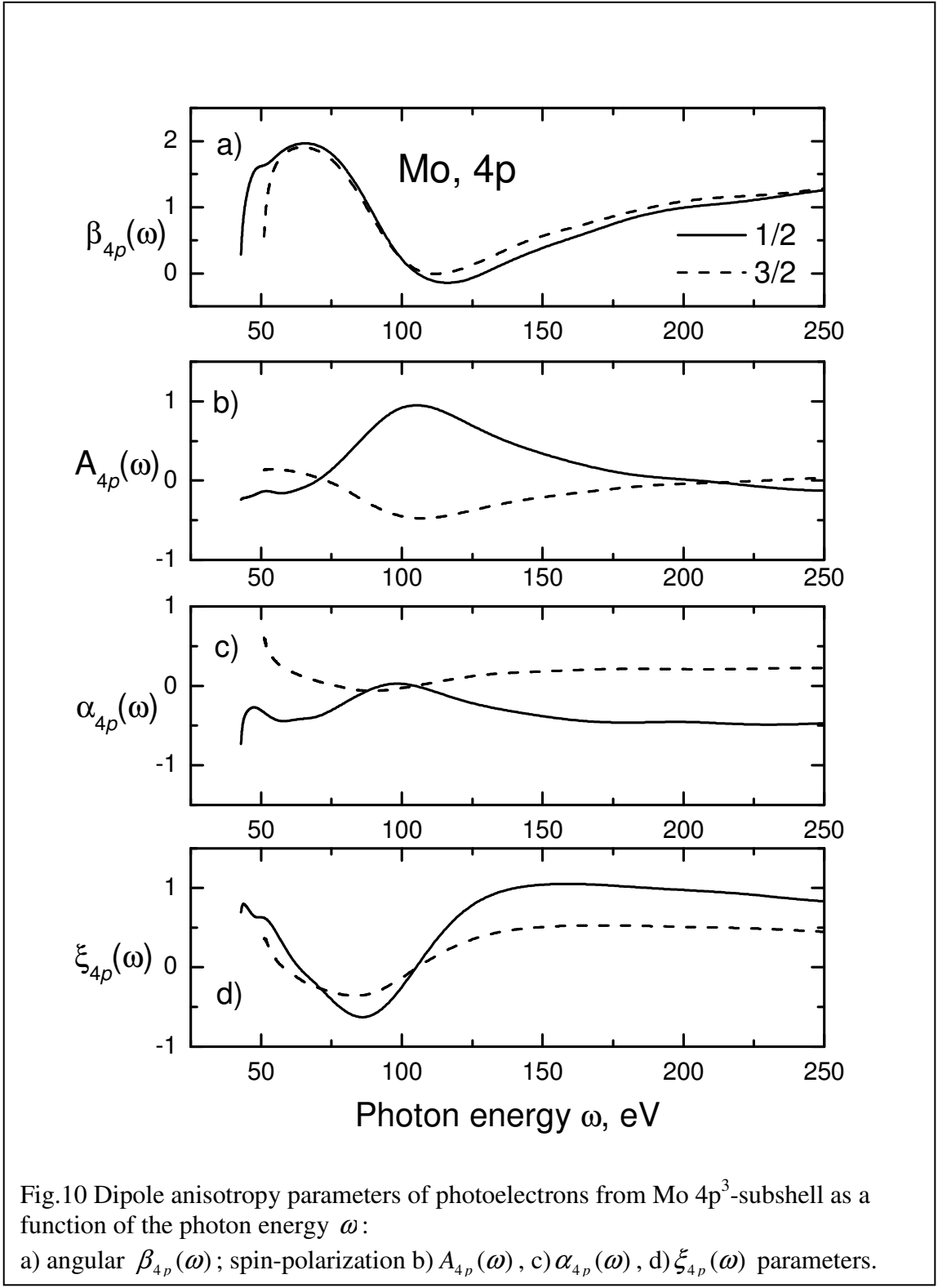

Fig.10 Dipole anisotropy parameters of photoelectrons from Mo $4p^3$-subshell as a function of the photon energy $\omega$:
a) angular $\beta_{4p}(\omega)$; spin-polarization b) $A_{4p}(\omega)$, c) $\alpha_{4p}(\omega)$, d) $\xi_{4p}(\omega)$ parameters.

Mo belongs to the same group as Cr, but their outer electrons $p$ parameters are considerably different, due to difference in principal quantum numbers.

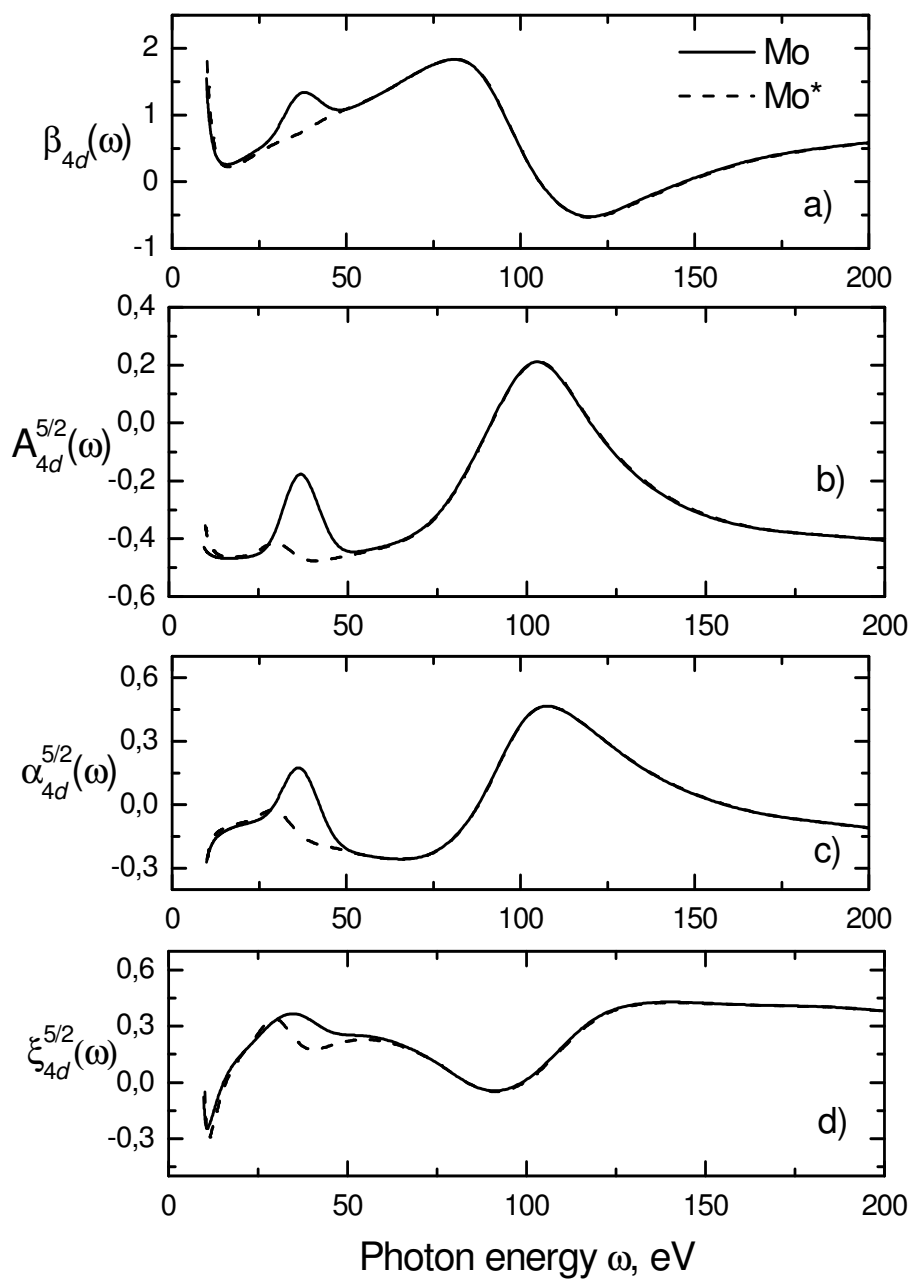

Fig.11 Dipole anisotropy parameters of photoelectrons from Mo and Mo* $4d^5$-subshell as a function of the photon energy $\omega$:
a) angular $\beta_{4d}(\omega)$; spin-polarization b) $A_{4d}^{5/2}(\omega)$, c) $\alpha_{4d}^{5/2}(\omega)$, d) $\xi_{4d}^{5/2}(\omega)$ parameters.

Note that all parameters of excited Mo* have a remarkable peculiarity at about 35 eV and in general their structure is more pronounced than in $Cr^*$.

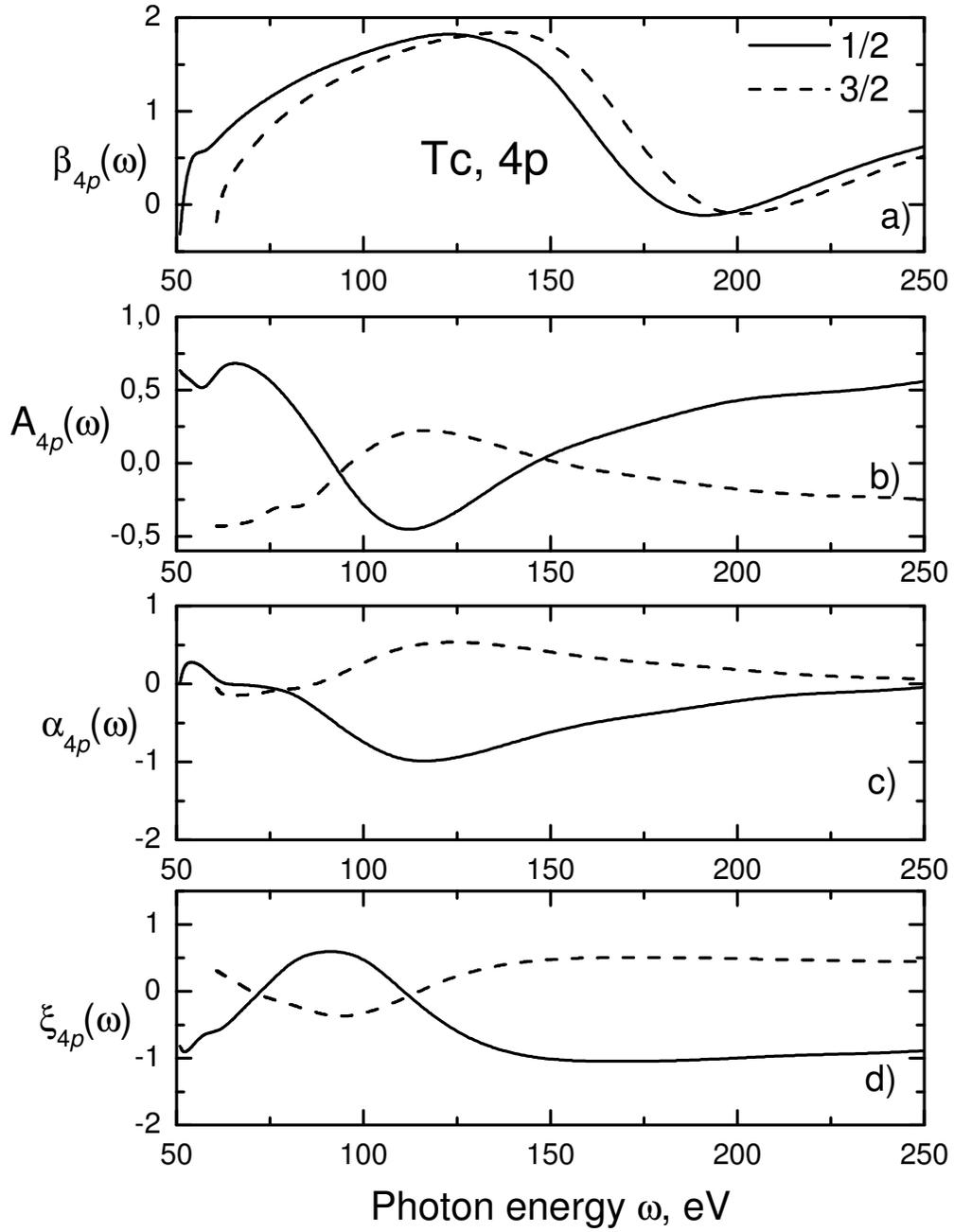

Fig.12 Dipole anisotropy parameters of photoelectrons from Tc $4p^3$-subshell as a function of the photon energy $\omega$:
a) angular $\beta_{4p}(\omega)$; spin-polarization b) $A_{4p}(\omega)$, c) $\alpha_{4p}(\omega)$, d) $\xi_{4p}(\omega)$ parameters.

The parameters of Tc have some essential common features to that of Mn. Particularly close are $\beta$ and $\alpha$ parameters. The similarity is stronger for $j=3/2$ than for $j=1/2$.

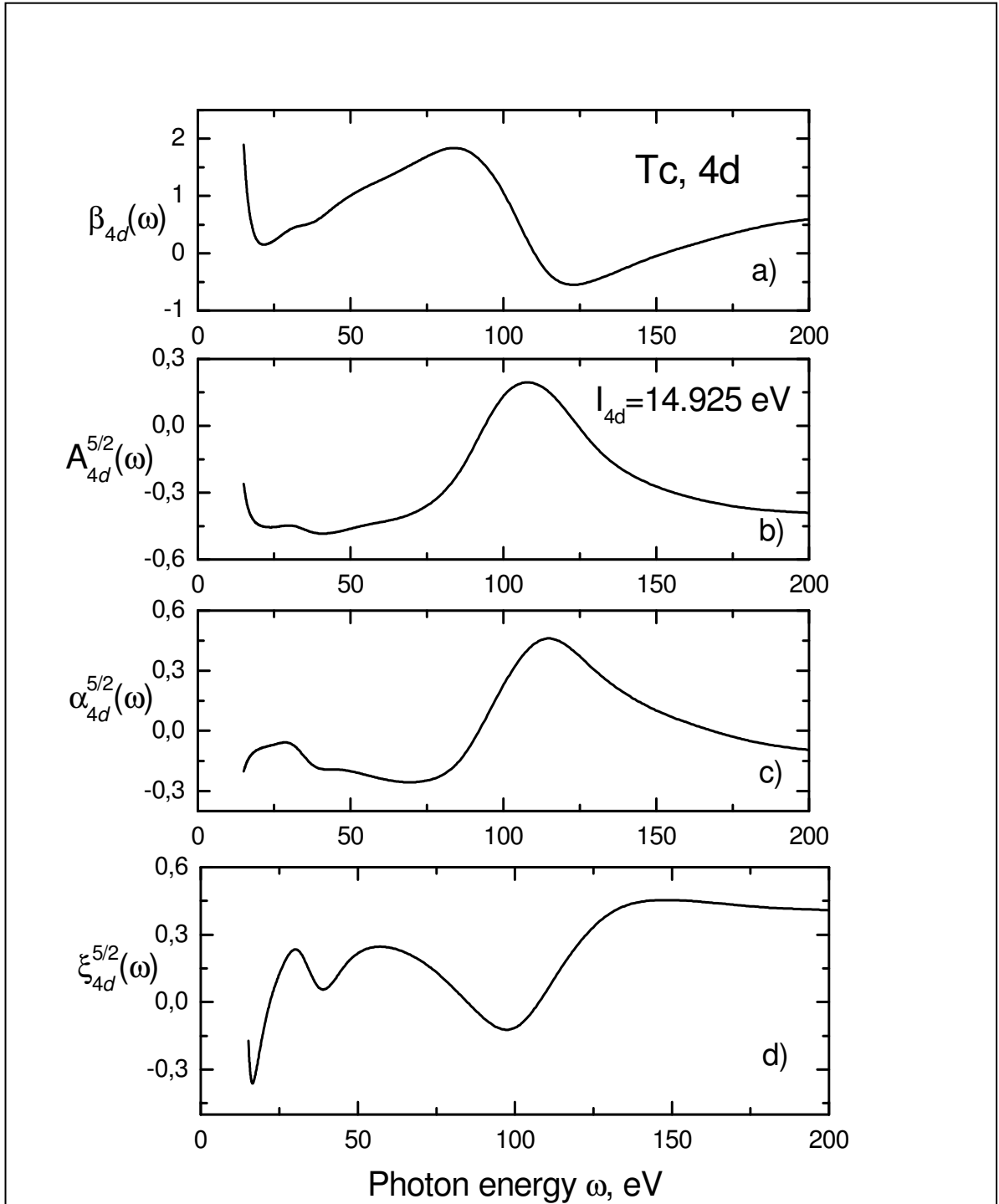

Fig. 13 Dipole anisotropy parameters of photoelectrons from Tc $4d^5$-subshell as a function of the photon energy $\omega$:
a) angular $\beta_{4d}(\omega)$; spin-polarization b) $A_{4d}^{5/2}(\omega)$, c) $\alpha_{4d}^{5/2}(\omega)$, d) $\xi_{4d}^{5/2}(\omega)$ parameters.

Tc spin-polarization parameters are remarkably similar to those of Mn. As to $\beta$, it has in Tc an extra variation with an asymmetric maximum at about 85 eV.

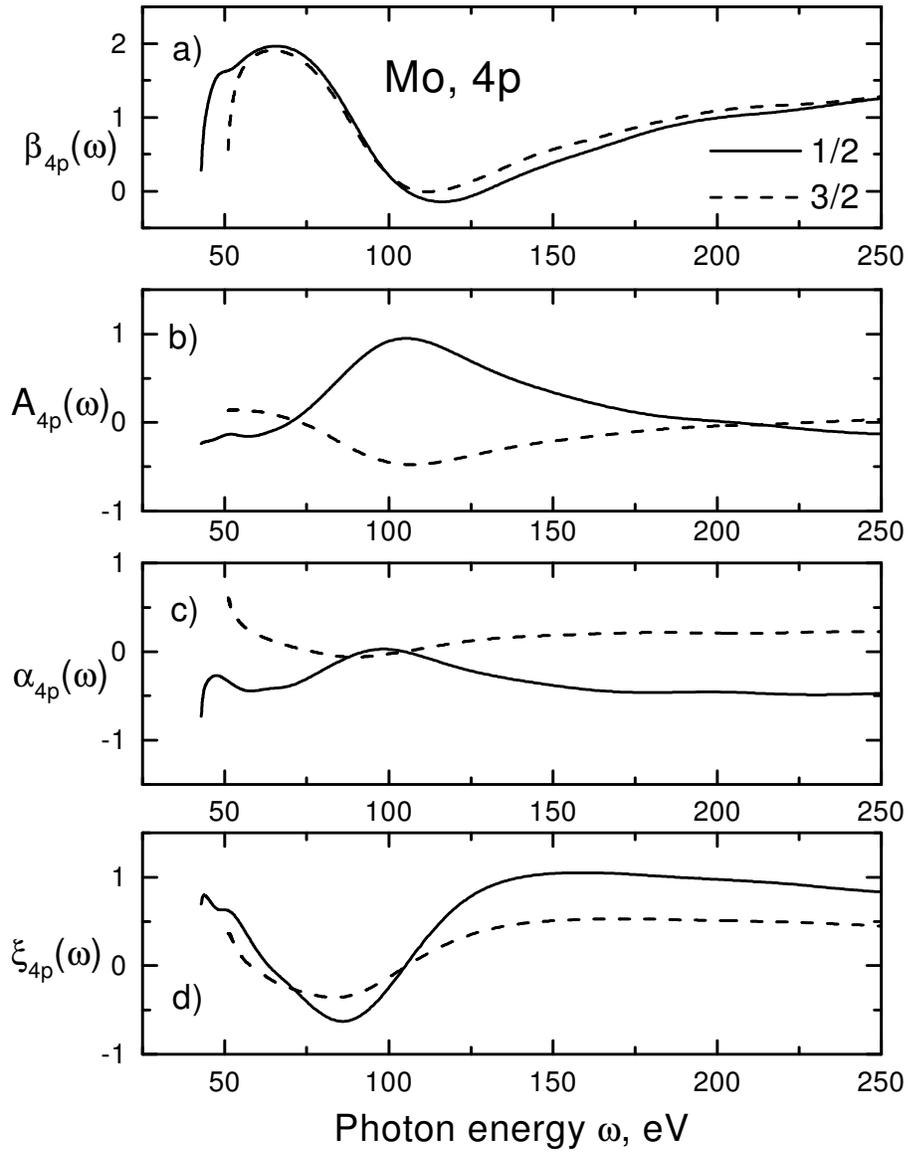

Fig. 14 Dipole anisotropy parameters of photoelectrons from Sb $4d^5$-subshell as a function of the photon energy $\omega$:
a) angular $\beta_{4d}(\omega)$; spin-polarization b) $A_{4d}(\omega)$, c) $\alpha_{4d}(\omega)$, d) $\xi_{4d}(\omega)$ parameters.

Sb belongs to the same group as As, but has essentially different parameters. Of special interest is the structure at 150-170 eV, no traces of which exists in

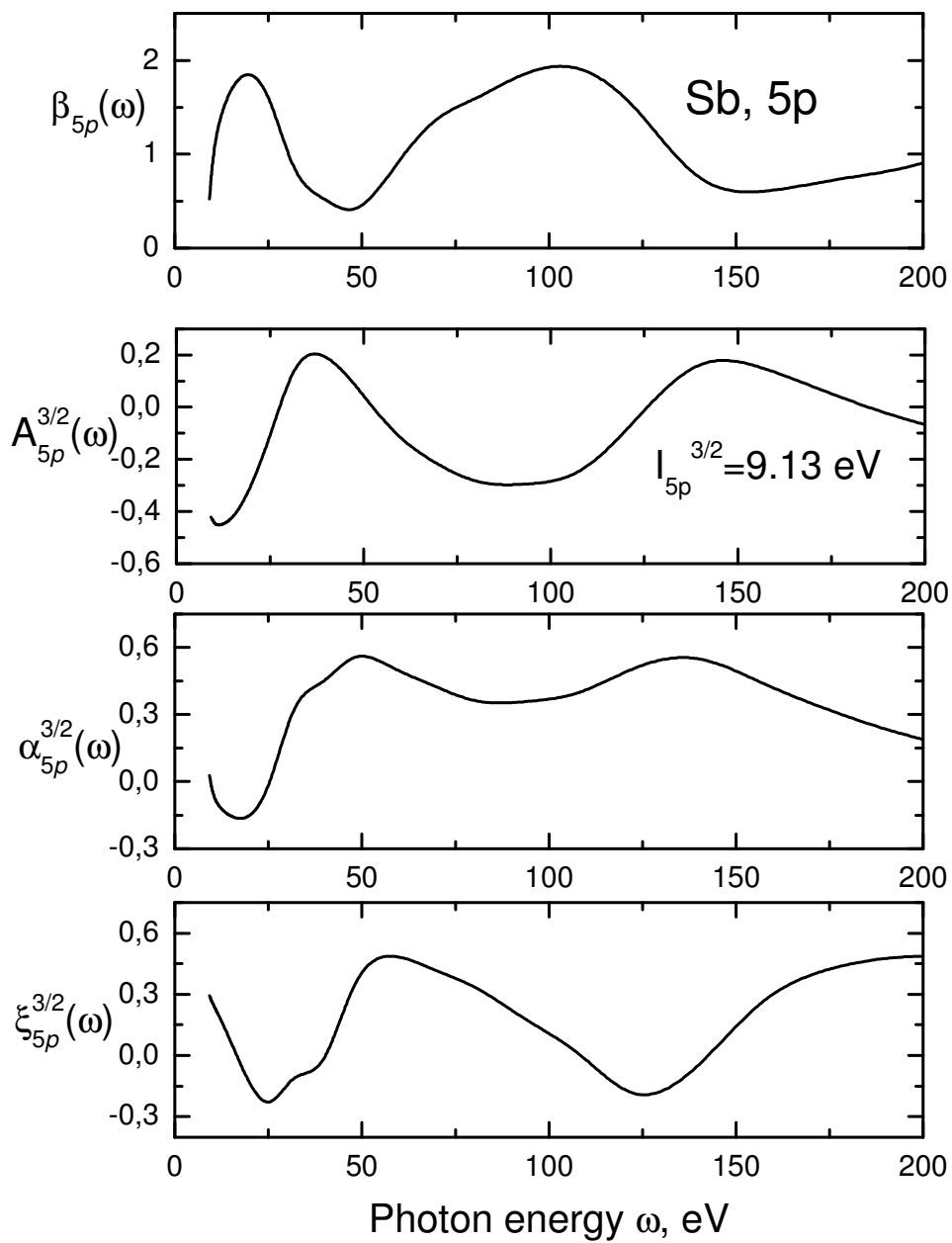

Fig. 15 Dipole anisotropy parameters of photoelectrons from Sb $5p^3$-subshell as a function of the photon energy $\omega$:
a) angular $\beta_{5p}(\omega)$; spin-polarization b) $A_{4p}(\omega)$, c) $\alpha_{4p}(\omega)$, d) $\xi_{4p}(\omega)$ parameters.

The parameters of Sb 5p electrons are complicated oscillating functions that have not too much in common with similar data on As.

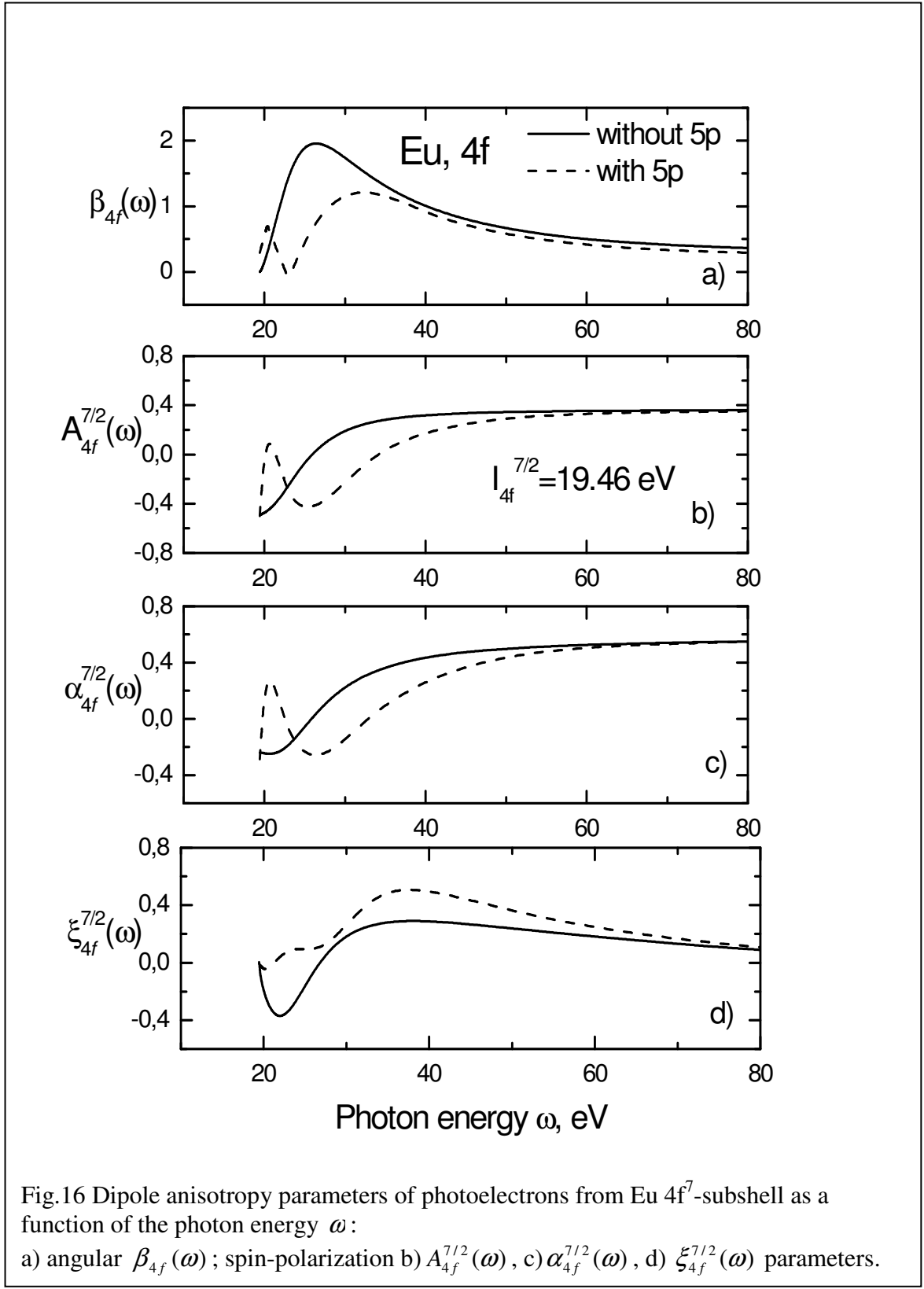

Fig.16 Dipole anisotropy parameters of photoelectrons from Eu $4f^7$-subshell as a function of the photon energy $\omega$:
a) angular $\beta_{4f}(\omega)$; spin-polarization b) $A_{4f}^{7/2}(\omega)$, c) $\alpha_{4f}^{7/2}(\omega)$, d) $\xi_{4f}^{7/2}(\omega)$ parameters.
21

We use Eu not only to show relatively simple behavior of considered parameters as functions of photon energy, but also to demonstrate the prominent role of 5*p* electrons in shaping the angular distribution and spin polarization of photoelectrons from the semi-filled 4*f* electrons.

Entirely, in semi-filled subshells and their neighbors we see a whole variety of structure that deserves experimental investigation.

**Acknowledgments**

MYaA is grateful to the Israeli Science Foundation, grant 174/03 - for the financial assistance to this research.